\documentclass[showpacs]{revtex4}

\usepackage{amsmath,latexsym,amssymb,amsfonts,mathbbol,amsthm,bm,citesort}

\renewcommand{\v}[1]{\mathbf{#1}}

\newcommand{\EE}[0]{\mathbb{E}}
\newcommand{\RR}[0]{\mathbb{R}}

\newcommand{\ZZ}[0]{\mathbb{Z}}

\newcommand{\NN}[0]{\mathbb{N}}

\newlength{\minus}
\newcommand{\ms}[0]{\hspace{\minus}}
\DeclareMathOperator{\diag}{diag}

\theoremstyle{plain}
\newtheorem{theorem}[subsection]{Theorem}

\begin{document}

\title{Semigroup extensions of isometry groups of compactified spacetimes}
\author{Hanno Hammer}
\email{H.Hammer@umist.ac.uk}
\affiliation{Department of Mathematics, UMIST \\
PO Box 88, Manchester M60 1QD, UK}

\date{\today}

\begin{abstract}
We investigate the possibility of semigroup extensions of the isometry
group of an identification space, in particular, of a compactified
spacetime arising from an identification map $p: \RR^n_t \rightarrow
\RR^n_t / \Gamma$, where $\RR^n_t$ is a flat pseudo-Euclidean covering
space and $\Gamma$ is a discrete group of primitive lattice
translations on this space. We show that the conditions under which
such an extension is possible are related to the index of the metric
on the subvector space spanned by the lattice vectors: If this
restricted metric is Euclidean, no extensions are
possible. Furthermore, we provide an explicit example of a semigroup
extension of the isometry group of the identification space obtained
by compactifying a Lorentzian spacetime over a lattice which contains
a lightlike basis vector. The extension of the isometry group is shown
to be isomorphic to the semigroup $(\ZZ^{\times},\cdot)$, i.e. the set
of nonzero integers with multiplication as composition and $1$ as unit
element. A theorem is proven which illustrates that such an extension
is obstructed whenever the metric on the covering spacetime is
Euclidean.
\end{abstract}

\pacs{22 E 70, 22 E 99.}
\maketitle


\section{Introduction}

In this work we examine the structure of the isometry group of orbit
spaces which are obtained from the action of a discrete group on a
flat pseudo-Euclidean covering space. The discrete group will be
realized as the set of primitive lattice transformations of a lattice
in the covering space which contains a lightlike lattice vector. In
this case, the restriction of the pseudo-Euclidean metric to the
lattice is no longer (positive or negative) definite. This gives rise
to the possibility of having lattice-preserving transformations in the
overall isometry group that are injective, but no longer surjective on
the lattice; in other words, their inverses do not preserve the
lattice. The set of all these transformations therefore will no longer
be a group, but only a semigroup. Since it is precisely the
lattice--preserving transformations that descend to the quotient of
the covering spacetime over the discrete group of primitive lattice
translations, these semigroup elements constitute an extension of the
isometry group of the compactified space which act non-invertibly on
the identification space. We present in detail the case of the
compactification of a Lorentzian spacetime over a lightlike lattice,
where it is shown that the non-invertible elements wind the lightlike
circle $k$ times around itself. We also present a theorem which
reveals that the possibility of semigroup extensions arises whenever
the restriction of the metric of the covering spacetime to the
subspace spanned by the lattice vectors is non-Euclidean. As a
consequence, semigroup extensions of isometry groups can be expected
in every scenario involving a torus compactification of a
higher-dimensional spacetime along a lightlike direction.

The plan of the paper is as follows: In section \ref{Sc.1} we provide
some background on orbit spaces and the associated fibre-preserving
sets. In section \ref{Sc.3} we study identifications over lattices and
examine a condition which obstructs nontrivial extensions of the
isometry group on the identification space. In section \ref{Sc.4} we
give an explicit example of a non-trivial extension, provided by
compactifying a flat Lorentzian spacetime over a lightlike lattice. It
will be shown that the extension of the isometry group in this case is
isomorphic to the semigroup $(\ZZ^{\times},\cdot)$ of nonzero integers
with multiplication as composition and $1$ as unit element.

\section{Orbit spaces and normalizing sets \label{Sc.1}}

Assume that a group $G$ has a left action on a topological space $X$
such that the map $G\times X\ni (g,x)\mapsto gx\in X$ is a
homeomorphism. When a discrete subgroup $\Gamma\subset G$ acts
properly discontinuously and freely on $X$, then the natural
projection $p:X\rightarrow X/\Gamma$ of $X$ onto the space of orbits,
$X/\Gamma $, can be made into a covering map, and $X$ becomes a
covering space of $X/\Gamma $
(e.g. \cite{Brown,Fulton,Jaehnich,Massey}). More specifically, if $X=M$
is a connected pseudo-Riemannian manifold with a metric $\eta $, and
$G=I(M)$ is the group of isometries of $M$, so that $\Gamma$ is a
discrete subgroup of isometries acting on $M$, then there is a unique
way to make the quotient $M/\Gamma$ a pseudo-Riemannian manifold
(e.g. \cite {Wolf,Poor,SagleWalde,Warner,ONeill}); in this construction
one stipulates that the projection $p$ be a local isometry, which
determines the metric on $M/\Gamma$. In such a case, we call
$p:M\rightarrow M/\Gamma $ a pseudo-Riemannian covering.

In any case the quotient $p:X\rightarrow X/\Gamma$ can be regarded as
a fibre bundle with bundle space $X$, base $X/\Gamma$, and $\Gamma$ as
structure group, the fibre over $m\in X/\Gamma$ being the orbit of any
element $x\in p^{-1}(m)$ under $\Gamma $, i.e. $p^{-1}(m)=\Gamma
x=\left\{\gamma x\mid \gamma \in \Gamma\right\}$. If $g\in G$ induces
the homeomorphism $x\mapsto gx$ of $X$ [or an isometry of $M$], then
$g$ gives rise to a well-defined map $g_{\#}:X/\Gamma\rightarrow
X/\Gamma $, defined by
\begin{equation}
\label{map1}
 g_{\#}(\Gamma x) \equiv \Gamma(gx) \quad,
\end{equation}
on the quotient space {\bf only} when $g$ preserves all fibres,
i.e. when $g\left(\Gamma x\right)\subset\Gamma(gx)$ for all $x\in
X$. This is equivalent to saying that $g\Gamma g^{-1}\subset
\Gamma$. If this relation is replaced by the stronger condition
$g\Gamma g^{-1}=\Gamma$, then $g$ is an element of the {\it
normalizer} \cite{Lang} $N\left(\Gamma\right)$ of $\Gamma$ in $G$,
where
\begin{equation}
\label{pp3fo1}
N\left(\Gamma\right)=\left\{g\in G\mid g\Gamma g^{-1}=\Gamma \right\}
\quad .
\end{equation}
The normalizer is a group by construction. It contains all
fibre-preserving elements $g$ of $G$ such that $g^{-1}$ is
fibre-preserving as well. In particular, it contains the group
$\Gamma$, which acts trivially on the quotient space; this means, that
for any $\gamma\in \Gamma$, the induced map
$\gamma_{\#}:X/\Gamma\rightarrow X/\Gamma$ is the identity on
$X/\Gamma $. This follows, since the action of $\gamma_{\#}$ on the
orbit $\Gamma x$, say, is defined to be $\gamma_{\#}\left(\Gamma
x\right):=\Gamma\left(\gamma x\right) =\Gamma x$, where the last
equality holds, since $\Gamma$ is a group.

In this work we are interested in relaxing the equality in the
condition defining $N\left(\Gamma\right)$; to this end we introduce
what we wish to call the {\it extended normalizer}, denoted by
$eN\left(\Gamma\right)$, through
\begin{equation}
\label{pp3fo2}
eN\left(\Gamma\right):=\left\{g\in G\mid g\Gamma g^{-1}\subset \Gamma
\right\} \quad .
\end{equation}
The elements $g\in G$ which give rise to well-defined maps $g_{\#}$ on
$ X/\Gamma$ are therefore precisely the elements of the extended
normalizer $eN\left(\Gamma\right)$, as we have seen in the discussion
above. Such elements $g$ are said to {\it descend to the quotient
space} $X/G$. Hence $eN\left(\Gamma\right)$ contains all
homeomorphisms of $X$ [isometries of $M$] that descend to the quotient
space $X/\Gamma$ [$M/\Gamma$]; the normalizer $N\left(\Gamma \right)$,
on the other hand, contains all those $g$ for which $g^{-1}$ descends
to the quotient as well. Thus, $N\left(\Gamma\right)$ is the group of
all $g$ which descend to {\bf invertible} maps $g_{\#}$
[homeomorphisms; isometries] on the quotient space. In the case of a
semi-Riemannian manifold $M$, for which the group $G$ is the isometry
group $I\left( M\right) $, the normalizer $N\left(\Gamma\right)$
therefore contains all isometries of the quotient space, the only
point being that the action of $N\left(\Gamma\right)$ is not
effective, since $\Gamma\subset N\left(\Gamma\right)$ acts trivially
on $M/\Gamma$. However, $\Gamma$ is a normal subgroup of
$N\left(\Gamma\right)$ by construction, so that the quotient
$N\left(\Gamma\right)/\Gamma$ is a group again, which is now seen to
act effectively on $M/\Gamma$, and the isometries of $M/\Gamma$ which
descend from isometries of $M$ are in a 1--1 relation to elements of
this group. Thus, denoting the isometry group of the quotient space
$M/\Gamma$ as $I\left(M/\Gamma\right)$, we have the well-known result
that
\begin{equation}
\label{pp3fo3}
 I (M/\Gamma)=N (\Gamma) / \Gamma \quad .
\end{equation}

Now we turn to the extended normalizer. For an element $g\in
eN\left(\Gamma\right)$, but $g\not\in N\left(\Gamma\right)$, the
induced map $g_{\#}$ is no longer injective on $X/\Gamma$: To see this
we observe that now the inclusion in definition (\ref{pp3fo2}) is
proper, i.e. $g\Gamma g^{-1} \subsetneqq\Gamma $. It
follows that a $\gamma' \in \Gamma$ exists for which
\begin{equation}
\label{hilf2}
g \gamma g^{-1} \neq \gamma' \quad \text{for all $\gamma \in \Gamma$.}
\end{equation}
Take an arbitrary $x \in M$; we claim that
\begin{equation}
\label{hilf3}
 g \Gamma g^{-1} x \subsetneqq \Gamma x \quad.
\end{equation}
To see this, assume to the contrary that the sets $g \Gamma g^{-1} x$
and $\Gamma x$ coincide; then $\gamma_1, \gamma' \in \Gamma$ exist
for which $g \gamma_1 g^{-1} x = \gamma' x$; since $g$ is an element
of the extended normalizer $eN(\Gamma)$, $g \gamma_1 g^{-1} \in
\Gamma$, i.e., $g \gamma_1 g^{-1} = \gamma_2$, say. It follows that
$\gamma_2 x = \gamma' x$ or $\gamma_2^{-1} \gamma' x = x$. The element
$\gamma_2^{-1} \gamma'$ belongs to $\Gamma$, which, by assumption,
acts freely. Free action implies that if a group element has a fixed
point then it must be the unit element, implying that $\gamma' =
\gamma_2 = g \gamma_1 g^{-1}$, which contradicts (\ref{hilf2});
therefore, (\ref{hilf3}) must hold. Eq. (\ref{hilf3}) can be expressed
by saying that the orbit of $x$ is the image of the orbit of $g^{-1}
x$ under the action of the induced map $g_{\#}$, $g_{\#} (\Gamma
g^{-1} x) = \Gamma x$; but that the $g_{\#}$-image of the orbit
$\Gamma g^{-1} x$, regarded as a set, is properly contained in the
orbit of $x$. The latter statement means that a $\gamma' \in \Gamma$
exists, as in (\ref{hilf2}), such that $\gamma' x \neq g \gamma g^{-1}
x$ for all $\gamma$. It follows that $\gamma g^{-1} x \neq g^{-1}
\gamma' x$ for all $\gamma$, implying that the point $g^{-1} \gamma'
x$ is not contained in the orbit of the point $g^{-1} x$. Its own
orbit, $\Gamma g^{-1} \gamma' x$, is therefore distinct from the orbit
$\Gamma g^{-1} x$ of the point $g^{-1} x$, since two orbits either
coincide or are disjoint otherwise. However, the induced map $g_{\#}$
maps $\Gamma g^{-1} \gamma' x$ into
\begin{equation}
\label{hilf4}
 g_{\#} (\Gamma g^{-1} \gamma' x) = \Gamma( g g^{-1} \gamma' x) =
 \Gamma x \quad,
\end{equation}
from which it follows that $g_{\#}$ maps two distinct orbits into the
same orbit $\Gamma x$, expressing the fact that $g_{\#}$ is not
injective. In particular, if $g$ was an isometry of $M$, then $g_{\#}$
can no longer be an isometry on the quotient space, since it is not
invertible on the quotient space. From this fact, or directly from its
definition (\ref{pp3fo2}), it also follows that $eN(\Gamma)$ is a
semigroup.

In this work we will show that the extended normalizer naturally
emerges when we study identification spaces $M/\Gamma$, where $M =
({\mathbb{R}}^n,\eta)$ is flat $\mathbb{R}^n$ endowed with a symmetric
bilinear form $\eta$ with signature $(-t,+s)$, $t+s=n$; to such a
space we will also refer to as $M=\mathbb{R}_t^n$. The isometry group
$I\left(\mathbb{R}_t^n\right)$ of $\mathbb{R}_t^n$ is the semi-direct
product
\begin{equation}
\label{pp3not1}
 I\left(\mathbb{R}_t^n\right) = \EE(t,n-t) = \mathbb{R}^n \odot
O(t,n-t) \quad ,
\end{equation}
called {\it pseudo-Euclidean group} $\EE(t,n-t)$, where the
translational factor $\mathbb{R}^n$ is normal in
$\EE(t,n-t)$. Elements of $\EE(t,n-t)$ will be denoted by $(t,R)$ with
group law $(t,R)(t',R')=(Rt'+t,RR')$. The Lie algebra of the
pseudo-Euclidean group $\EE(t,n-t)$ will be denoted by
$\mathsf{eu}(t,n-t)$ henceforth.

For $t=1$, $\EE(1,n-1)$ is the Poincare group, and $\RR^n_1$ is a
Lorentzian manifold, with metric $\eta = \diag(-1,1, \ldots, 1)$. In
this case we call elements $v \in \RR^n_1$ timelike, lightlike,
spacelike if $\eta(v,v) <0$, $=0$, $>0$, respectively.

The identification group $\Gamma$ will be realized as a discrete group
of translations in $M$, the elements being in 1--1 correspondence with
the points of a lattice $\mathsf{\mathsf{lat}} \subset
\mathbb{R}_t^n$, which is regarded as a subset of $\mathbb{R}_t^n$. We
will show explicitly that in the Lorentzian case, the fact that the
identity component $SO_0(1,n-1) \subset O(1,n-1)$ of the Lorentz group
is no longer compact gives rise to a natural extension of the isometry
group $N(\Gamma) / \Gamma$ of the quotient $M / \Gamma$ to the set
$eN(\Gamma) / \Gamma$, provided that the $\mathbb{R}$-linear envelope
of the lattice is a lightlike subvector space. This will be compared
to the orthogonal case, and it will be shown in Theorem
[\ref{Bedingung}] that the compactness of $SO(n)$ obstructs such an
extension. That is probably why such extensions have not been studied
in the field of crystallography in the past.

\section{Identifications over a lightlike lattice \label{Sc.3}}

As mentioned above, we shall study quotient spaces of flat covering
spacetimes over a group $\Gamma$ of primitive lattice
transformations. If the $\RR$-linear span of the lattice vectors has
(real) dimension $m$, say, the resulting identification space
$M/\Gamma$ is homeomorphic to a product manifold $\RR^{n-m}\times
T^m$, where $T^m$ denotes an $m$-dimensional torus. This space
inherits the metric from the covering manifold $\RR^n_t$, since the
metric is a local object, but the identification changes only the
global topology. Thus $M/\Gamma$ is again a semi-Riemannian
manifold. Whereas the isometry group of the covering space $\RR^n_t$
is $\EE(t,n-t)$, the isometry group of the ``compactified'' space
$M/\Gamma$ is obtained from the normalizer $N(\Gamma)$ of the group
$\Gamma$ in $\EE(t,n-t)$ according to formula (\ref{pp3fo3}).

In this work we restrict attention to lattices that contain the origin
$0\in\mathbb{R}_t^n$ as a lattice point, which suffices for our
purposes. Let $1\le m\le n$, let $\underline{u}\equiv(u_1,\ldots
,u_m)$ be a set of $m$ linearly independent vectors in
$\mathbb{R}_t^n$; then the $\mathbb{Z}$-linear span of
$\underline{u}$,
\begin{equation}
\label{pp3fo4}
 \mathsf{\mathsf{lat}} \equiv \sum_{i=1}^m \mathbb{Z} \cdot u_i=\left\{
\sum_{i=1}^m z_i \cdot u_i \mid z_i \in \mathbb{Z} \right\} \quad ,
\end{equation}
is called the set of lattice points with respect to
$\underline{u}$. Elements of $\mathsf{\mathsf{lat}}$ are regarded as
points in $\mathbb{R}_t^n$ as well as vectors on the tangent space
$T\mathbb{R}_t^n$ of $\mathbb{R}_t^n$. Let $[\mathsf{\mathsf{lat}}]
\equiv[\underline{u}]_{\mathbb{R}}$ denote the $\mathbb{R}$-linear
span of $\mathsf{\mathsf{lat}}$.

We recall that the {\it index} $ind(W)$ of a vector subspace
$W\subset\mathbb{R}_t^n$ is the maximum in the set of all integers
that are the dimensions of $\mathbb{R}$-vector subspaces $W'\subset W$
on which the restriction of the metric $\left.\eta\right|W'$ is
negative definite, see e.g. \cite{ONeill}. Hence $0\le ind(W)\le m$,
and $ind(W)=0$ if and only if $\left.\eta\right|W$ is positive
definite. In the Lorentzian case, i.e. $M=\mathbb{R}_1^n$, we call $W$
timelike $\Leftrightarrow$ $\left.\eta\right|W$ is nondegenerate, and
$ind(W)=1$; $W$ lightlike $\Leftrightarrow$ $\left.\eta\right|W$
degenerate, and $W$ contains a $1$-dimensional lightlike vector
subspace, but no timelike vector; and $W$ spacelike $\Leftrightarrow$
$\left.\eta\right|W$ is positive definite and hence $ind(W)=0$.

We define the {\it index of the lattice} as the index of its
$\mathbb{R}$-linear span $[\mathsf{\mathsf{lat}}]$,
$ind(\mathsf{lat})\equiv ind([\mathsf{\mathsf{lat}}])$.

Let $T_{\mathsf{lat}}\subset \EE(t,n-t)$ be the subgroup of all
translations in $ \EE(t,n-t)$ through elements of
$\mathsf{\mathsf{lat}}$,
\begin{equation}
\label{pp3fo5}
 T_{\mathsf{lat}}=\left\{(t_z,1)\in \EE(t,n-t) \mid t_z\in
 \mathsf{lat} \right\} \quad .
\end{equation}
Elements $(t_z,1)$ of $T_{\mathsf{lat}}$ are called {\it primitive
translations}. $T_{\mathsf{lat}}$ is taken as the discrete group
$\Gamma$ of identification maps, which give rise to the identification
space $p:\RR^n_t \rightarrow \RR^n_t/\Gamma$ under study.

We now examine the normalizer and extended normalizer of the
identification group: For an element $(t,R)\in\EE(t,n-t)$ to be in the
extended normalizer $eN(\Gamma)$ of $\Gamma$, the condition
\begin{equation}
(t,R)(t_z,1)(t,R)^{-1} = (Rt_z,1) \in T_{\mathsf{lat}}
\label{Beding1}
\end{equation}
must be satisfied for lattice vectors $t_z$. In other words, $Rt_z\in
\mathsf{lat}$, which means that the pseudo-orthogonal transformation
$R$ must preserve the lattice $\mathsf{\mathsf{lat}}$, $R \mathsf{lat}
\subset \mathsf{lat}$. For an element $(t,R)$ to be in the normalizer,
$(t,R)^{-1}$ must be in the normalizer as well, implying $R^{-1}
\mathsf{lat} \subset \mathsf{lat}$, so altogether $R \mathsf{lat} =
\mathsf{lat}$. The elements $R$ occuring in the normalizer therefore
naturally form a subgroup $G_{\mathsf{lat}}$ of $O(t,n-t)$; on the
other hand, the elements $R$ occuring in the extended normalizer form
a semigroup $eG_{\mathsf{lat}}\supset G_{\mathsf{lat}}$. Furthermore,
no condition on the translations $t$ in $(t,R)$ arises, hence all
translations occur in the [extended] normalizer. Thus, the [extended]
normalizer has the structure of a semi-direct [semi-]group
\begin{align}
N\left( \Gamma \right) & = \mathbb{R}^n\odot G_{\mathsf{lat}} \quad , 
\label{Norm1}\\
eN\left( \Gamma \right) & = \mathbb{R}^n\odot eG_{\mathsf{lat}} \quad , 
\tag{\ref{Norm1}'}
\end{align}
where $\RR^n$ refers to the subgroup of all translations in
$\EE(t,n-t)$.

We now present a condition under which $eG_{\mathsf{lat}}$ coincides with
$G_{\mathsf{lat}}$:

\begin{theorem}[Condition] \label{Bedingung}
If\, $ind(\mathsf{lat}) = 0$ or $ind(\mathsf{lat}) = m$ (i.e. minimal
or maximal), then $eG_{\mathsf{lat}} = G_{\mathsf{lat}}$.
\end{theorem}

{\it Proof:}

We first assume that $ind(W) = 0$, i.e. $\left. \eta \right|W$ is
positive definite. Let $R\in eG_{\mathsf{lat}} \subset O(t,n-t)$. Since $R$
preserves $\mathsf{\mathsf{lat}}$, it also preserves its $\mathbb{R}$--linear envelope
$W$, i.e. $R W\subset W$. Let $x,y\in W$ arbitrary, then $
R x,R y\in W$. 
Since $\eta(Rx,Ry) =\eta(x,y)$, the restriction $\left. R \right| W$
of $R $ to the subvector space $W$ preserves the bilinear form
$\left. \eta \right| W$ on this space. But $\left. \eta \right| W$ is
positive definite by assumption of $ind(W)=0$, hence $\left. R \right|
W\in O\left( W\right) $, where $ O\left( W\right) $ denotes the
orthogonal group of $W$.

Now we assume that $R$ has the property 
\begin{equation}
\label{pp3fo14}
R \in eG_{\mathsf{lat}} ,\quad \text{but\quad }R \not \in
G_{\mathsf{lat}} \quad ,
\end{equation}
in other words, $R \mathsf{lat} \subsetneqq \mathsf{lat}$. This means
that the restriction $\left. R \right| \mathsf{lat}$ is not
surjective. Hence
\begin{equation}
\label{hence1}
 \exists x \in \mathsf{lat} : R u \neq x \quad \text{for all $u \in
 \mathsf{lat}$} \quad.
\end{equation}
$x$ cannot be zero, since $0\in \mathsf{lat}$ , and $R $ is
linear. Hence $r\equiv \left\| x\right\| >0$, where $ \left\|x\right\|
=\sqrt{\eta \left( x,x\right) }$ denotes the Euclidean norm on
$W$. Now let $S_{m-1}$ be the $(m-1)$-dimensional sphere \underline{in
$W$}, centered at $0$. Consider the intersection $\mathsf{sct} =
\mathsf{lat} \cap r \cdot S_{m-1}$, where $r \cdot S_{m-1}$ is the
$\left(m-1\right)$-dimensional sphere with radius $r$ in $W$. Note
that $r \cdot S_{m-1}$ coincides with the orbit $O(m-1) \cdot x$ of
$x$ under the action of the orthogonal group $O(m-1)$, which is a
compact set on account of the fact that $O(m-1)$ is a compact Lie
group. From the compactness of $r \cdot S_{m-1}$ it follows that the
number of elements $\#\mathsf{sct}$ of $\mathsf{sct}$ is {\bf finite};
from $x \in \mathsf{sct}$ it follows that $\mathsf{sct}$ is not empty,
so $1 \le \#\mathsf{sct} < \infty $. Then
\begin{enumerate}
\item  
 $\left. R \right| W$ orthogonal \quad $\Rightarrow$ \quad $R \left(
 \mathsf{sct}\right) \subset r\cdot S_{n-1}$;
\item  
 $R$ lattice preserving \quad $\Rightarrow$ \quad $R \left(
 \mathsf{sct}\right) \subset \mathsf{lat}$;
\item  
 $R$ injective \quad $\Rightarrow$ \quad $\#R \left(
 \mathsf{sct}\right) = \#\left( \mathsf{sct}\right)$.
\end{enumerate}
The first two statements imply that $\left.R\right| W$ preserves
$\mathsf{sct}$, $\left( \left. R \right| W\right) \left(
\mathsf{sct}\right) \subset \mathsf{sct}$ ; from the third we deduce
that $\left( \left. R \right| W\right) \left( \mathsf{sct}\right) =
\mathsf{sct}$, since the set $\mathsf{sct}$ is finite. But this says
that all elements of $\mathsf{sct}$ are in the image of $\left(
\left. R \right| W\right) $, hence $x=\left( \left.  R \right|
W\right) \left( x^{\prime }\right) $ for some $x^{\prime }\in
\mathsf{sct}$, which is a contradiction to (\ref{hence1}). This
implies that our initial assumption (\ref{pp3fo14}) concerning $R $
was wrong.

Now assume that $ind\left(\mathsf{lat}\right) $ is maximal. Then
$\left. \eta \right| W$ is negative definite, but the argument given
above clearly still applies, since $O(0,m-1) \simeq O(m-1,0)$, and
the only point in the proof was the compactness of the
$O(m-1)$-orbits. This completes our proof.  {\hfill$\blacksquare$}

We see that the structure of the proof relies on the compactness of
orbits $O\cdot x$ of $x$ under the orthogonal group, which, in turn,
comes from the fact that the orthogonal groups $O$ are compact. If the
metric restricted to $[\mathsf{\mathsf{lat}}]$ were pseudo-Euclidean
instead, we could have non-compact orbits, related to the
non-compactness of the groups $O(t,n-t)$. In this case we expect the
possibility that $G_{\mathsf{lat}} \subsetneqq eG_{\mathsf{lat}}$ to
arise. An explicit example of this situation will be given now.

\section{Compactification of a Lorentzian spacetime} \label{Sc.4}

The covering spacetime is chosen to be $M=\RR^n_1$, i.e. an
$n$-dimensional Lorentzian spacetime with metric $\eta =\diag(-1,1,
\ldots, 1)$. The lattice $\underline{u}$ is constructed as follows:
Let $\left({\bf e}_0,{\bf e}_1,\ldots,{\bf e}_{n-1}\right)$ denote the
canonical basis of $\mathbb{R}^n_1$, where ${\bf e}_0$ is the ``time
direction'' with $\eta({\bf e}_0,{\bf e}_0)=-1$. We define ${\bf
e}_{\pm}\equiv\frac{1}{\sqrt{2}}({\bf e}_0\pm{\bf e}_1)$, so the new
basis vectors ${\bf e}_{\pm}$ are both lightlike. We split the new
basis according to
\begin{equation}
({\bf e}_+,{\bf e}_-,{\bf e}_2,\ldots,{\bf e}_{n-m},{\bf e}_{n-m+1},
\ldots,{\bf e}_{n-1}) \quad.
\label{split1}
\end{equation}
As lattice vectors we choose $\underline{u}\equiv({\bf e}_+,{\bf
e}_{n-m+1}, \ldots,{\bf e}_{n-1})$. Upon compactifying these
dimensions, i.e. taking the quotient of $\RR^n_1$ over the associated
group $T_{\mathsf{lat}}$, we obtain an $m$-dimensional torus $T^m$;
the remaining $n-m$ dimensions along $({\bf e}_-,{\bf e}_2,\ldots,{\bf
e}_{n-m})$ remain uncompactified, resulting in a product manifold
homeomorphic to $\RR^{n-m}\times T^m$.

We need to construct the lattice-preserving sets $G_{\mathsf{lat}}$
[$eG_{\mathsf{lat}}$] in order to obtain the [extended] normalizer
from formulae (\ref{Norm1}, \ref{Norm1}'). We proceed in several steps:
In step one we identify the set of Lorentz transformations on the
covering space $\RR_1^n$ which preserve the $\RR$-linear span of the
lattice vectors $\underline{u}$; thus, we have to identify those
generators of Lorentz transformations which map elements of the
lattice basis into linear combinations of these basis vectors. As will
be seen below, the set of Lorentz transformations so obtained will
form a Lie subgroup $G_1$ of $O(1,n-1)$. In step two we impose the
further condition that not only the $\RR$-linear span, but the
$\ZZ$-linear span of basis vectors $\underline{u}$, and hence the true
lattice points, shall be preserved. This restricts the group $G_1$
further down to a set $G_2$ which will be seen to contain a product of
a discrete group, a Lie group and a discrete semigroup, which is what
we wanted to show.

Let us proceed to step one: We seek  the Lorentz  transformations
$\Lambda$ which  map elements  of  $\underline{u}$ into real  linear
combinations  of  elements  of $\underline{u}$.  Now, $\underline{u}$
contains one lightlike, and $m-1$ spacelike basis vectors. Assume that
the  image $\Lambda\, \v{e}_+$ of  $\v{e}_+$ is  given  by a  linear
combination  $\Lambda\, \v{e}_+ = a\, \v{e}_+ + \sum_{i=n-m+1}^{n-1}
b_i\, \v{e}_i$. Since $\v{e}_+$ is lightlike and $\Lambda$ preserves
pseudo-Euclidean lengths we must  have  that  $\sum_{i=n-m+1}^{n-1}
b_i^2 = 0$, hence  $\Lambda \v{e}_+ = a\, \v{e}_+$ necessarily. As a
consequence, all admissible $\Lambda$ preserve the  one-dimensional
lightlike subvector space $\RR \cdot \v{e}_+$.  The task  of step one
is therefore performed  in two substeps: First  we identify the set of
Lorentz transformations which preserve  this subvector  space, and
secondly we  impose  the further restriction that the  real linear
span of all  basis vectors, not only $\v{e}_+$, be preserved.

We now identify the set of Lorentz transformations that preserve the
one-dimensional subvector space $\RR\cdot{\bf e}_+$. We do not wish to
compute this set in the full Lorentz group but, for sake of
simplicity, focus on the identity component $SO_0(1,n-1)$ of
$O(1,n-1)$ instead. The end result will still exhibit the semigroup
extension under consideration, and this is what we want to concentrate
on.

We thus seek the subset of orthochronous orientation-preserving
Lorentz transformations that preserve the light-like subvector space
$\RR \cdot \v{e}_+$. Since these matrices are in the identity
component of $O(1,n-1)$, they will be (products of) exponentials of
Lie algebra elements $L \in so(1,n-1)$ which map the subvector space
$\RR \cdot \v{e}_+$ into itself, including annihilation of
$\v{e}_+$. We use the following conventions for the matrices
$L_{\alpha\beta}$ of a standard basis of $so(1,n-1)$:
\begin{subequations}
\label{ots2}
\begin{align}
 L_{\alpha\beta} & = -L_{\beta\alpha} \quad, \quad 0 \le \alpha <
 \beta \le n-1 \quad, \label{ots2a} \\
 \left( L_{\alpha\beta}\right)_{\mu\nu} & = \eta_{\alpha\mu}\,
 \delta_{\beta\nu} - \eta_{\beta\mu}\, \delta_{\alpha\nu} \quad, \quad
 \forall\; \alpha, \beta \quad. \label{ots2b}
\end{align}
\end{subequations}
The Lie algebra of these matrices, realized by matrix commutators, is
\begin{equation}
\label{ots3}
 \left[ L_{\alpha\beta}, L_{\rho\sigma} \right] = -\eta_{\alpha\rho}\,
 L_{\beta\sigma} - \eta_{\beta\sigma}\, L_{\alpha\rho} +
 \eta_{\alpha\sigma}\, L_{\beta\rho} + \eta_{\beta\rho}\,
 L_{\alpha\sigma} \quad, \quad \forall\; \alpha, \beta \quad.
\end{equation}
The special Lorentz transformations we seek are generated by those
linear combinations of matrices (\ref{ots2}) which map the vector
$\v{e}_+$ onto a multiple of itself, including annihilation. It is
straightforward to find that these generators take the form
\begin{subequations}
\label{cond2}
\begin{align}
 & b \cdot L_{01} + \sum_{\alpha=2}^{n-1} v_{\alpha}\cdot U_{\alpha} +
 \sum_{2\le \alpha< \beta} \lambda_{\alpha\beta}\cdot L_{\alpha\beta}
 \quad, \label{cond2a} \\
 & U_{\alpha} = L_{0\alpha} - L_{1\alpha} \quad, \label{cond2b}
\end{align}
\end{subequations}
where $b \in \RR$, $\vec{v} \equiv (v_2, \ldots, v_{n-1}) \in
\RR^{n-2}$ and $\lambda_{\alpha\beta} \in \RR$, $2 \le \alpha < \beta
\le n-1$. The commutator algebra of the set $(L_{01}, U_{\alpha},
L_{\alpha \beta})$ is given as
\begin{subequations}
\label{cond3}
\begin{align}
 [ L_{01}, U_{\alpha}] & = - U_{\alpha} \quad, \label{cond3a} \\
 [ L_{01}, L_{\alpha \beta} ] & = \ms 0 \quad, \label{cond3b}
 \\
 [ U_{\alpha} , U_{\beta} ] & = \ms 0 \quad, \label{cond3c} \\
 [ L_{\alpha\beta} , U_{\gamma} ] & = - \eta_{\alpha \gamma}\,
 U_{\beta} + \eta_{\beta \gamma}\, U_{\alpha} \quad, \label{cond3d} \\
 [ L_{\alpha\beta}, L_{\rho\sigma} ] & = - \delta_{\alpha\rho}\,
 L_{\beta \sigma} - \delta_{\beta \sigma}\, L_{\alpha \rho} +
 \delta_{\alpha \sigma}\, L_{\beta \rho} + \delta_{\beta \rho}\,
 L_{\alpha \sigma} \quad. \label{cond3e}
\end{align}
\end{subequations}
We see that the commutator algebra of the generators (\ref{cond2})
closes, which is to be expected, since conceptually, the set of
Lorentz transformations preserving a one-dimensional subspace must
form a group. The set (\ref{cond2}) therefore forms a Lie algebra
$\hat{g}_1$, with associated Lie group $G_1$. The last three
relations, (\ref{cond3c}, \ref{cond3d}, \ref{cond3e}), express the
fact that the generators $(L_{\alpha\beta}, U_{\gamma})$, for $\alpha
\beta \gamma \ge 2$, span a subalgebra of $\hat{g}_1$ which is
isomorphic to the Euclidean algebra $\mathsf{eu}(n-2)$ in $n-2$ dimensions, the
$U_{\gamma}$ playing the part of translation generators. This
subalgebra is actually an ideal in $\hat{g}_1$, as follows from
(\ref{cond3a}, \ref{cond3b}). As a consequence, the Lie subgroup
$\EE(n-2)$ associated with the generators $(L_{\alpha \beta},
U_{\gamma})$ is normal in $G_1$; the latter therefore is homomorphic
to
\begin{equation}
\label{pp3fo23}
 G_1 \simeq \EE(n-2) \odot \RR^+ \quad,
\end{equation}
where $"\odot"$ denotes a semidirect product (accounting for the
normality of the $\EE$-factor), and $\RR^+$ denotes the group of all
real numbers $b$ with addition as group composition and zero as the
neutral element.

In a certain neighbourhood of the identity, the Lorentz
transformations generated by (\ref{cond2}) may be obtained in the form
\begin{equation}
\label{norco2}
\left.
\begin{gathered}
 \Lambda \left( b , \vec{v}, C\right) = S(\vec{v})\cdot R(C)
 \cdot B(b) = \\[8pt]
 = \left( 
\begin{array}{cccc}
 1+\frac{\vec{v}^2}2 &  -\frac{\vec{v}^2}2 & \vline & \vec{v}^T \\
 \frac{\vec{v}^2}2 & 1-\frac{\vec{v}^2}2 & \vline & \vec{v}^T \\[3pt] \hline
 \rule{0pt}{12pt} \vec{v} &  -\vec{v} & \vline &\Eins_{n-2}
\end{array}
\right) \cdot \left( 
\begin{array}{cccc}
1 & 0 & \vline & 0 \\ 
0 & 1 & \vline & 0 \\[3pt] \hline
0 & 0 & \vline & C 
\end{array}
\right) \cdot
\left( 
\begin{array}{cccc}
 \cosh b & \sinh b & \vline & 0 \\
 \sinh b & \cosh b & \vline & 0 \\[3pt] \hline
 \rule{0pt}{12pt} 0 & 0 & \vline & \Eins_{n-2}
\end{array}
\right) 
\quad,
\end{gathered}
\right.
\end{equation}
where $\vec{v}^2 = \sum_{\alpha=2}^{n-1} v_{\alpha}^2$, $C = C(\lambda
_{\alpha \beta})$ denotes elements of $SO(n-2)$, whilst $R(C)$
indicates rotations $C$ embedded in $SO_0(1,n-1)$, as in the second
matrix in (\ref{norco2}). In this representation, the parameters
$\vec{v}, \lambda_{\alpha\beta}, b$ serve as normal coordinates of the
second kind \cite{SagleWalde}.

The one-dimensional Lie subgroup represented by matrices $b \mapsto
B(b)$ can be expressed in a different way: To this end we substitute
$a = e^{b}$ in $\cosh b$ and $\sinh b$, with the result
that
\begin{equation}
\label{confgen}
 B(b) = B(\ln a) = 
\left( 
\begin{array}{cccc}
 \frac{a+\frac{1}{a}}{2} & \frac{a-\frac{1}{a}}{2} & \vline & 0 \\
 \frac{a-\frac{1}{a}}{2} & \frac{a+\frac{1}{a}}{2} & \vline & 0 \\[3pt] \hline
 \rule{0pt}{12pt} 0 & 0 & \vline & \Eins_{n-2}
\end{array}
\right) \quad.
\end{equation}
Since $\ln(a a') = \ln a + \ln a'$, these matrices now represent
$\RR_+^{\times}$, i.e. the group of all positive numbers with
multiplication as group composition and $1$ as the unit element. If
the same analysis as before is now repeated with respect to the full
Lorentz group $O(1,n-1)$, not only its connected component
$SO_+(1,n-1)$, we find that the group $\RR_+^{\times}$ is extended by
another component which comprises the negative numbers; as a
consequence, the matrices (\ref{confgen}) then represent the group of
nonvanishing real numbers $\RR^{\times}$ with multiplication and unit
element as before, according to $\RR^{\times} \ni a \mapsto \pm B(\ln
|a|)$.


So far we have obtained the Lie subgroup of the (connected component
of the) Lorentz group which preserves the one-dimensional subspace
$\RR\cdot{\bf e}_+$ spanned by the first lattice vector. We now
implement the task of substep two of step one and impose the further
condition that the whole real subspace spanned by all lattice vectors
be preserved. The matrices satisfying this stronger condition comprise
a subgroup $G_2\subset G_1$. We find that the Lie algebra $\hat{g}_2$
of $G_2$ may be generated by the same set of basis elements that span
$\hat{g}_1$, {\bf except} that the generators $L_{\alpha i}$, $2 \le
\alpha \le n-m$ and $n-m+1 \le i \le n-1$, now must be omitted. The
generators of $\hat{g}_2$ are therefore
\begin{subequations}
\label{geng2}
\begin{align}
 & L_{01} \quad, \label{geng2a} \\
 & (L_{\alpha\beta}, U_{\gamma}\, |\, 2 \le \alpha < \beta \le n-m ;
 \, 2 \le \gamma \le n-m) \quad, \label{geng2b} \\
 & (L_{ij}, U_k \, | \, n-m+1 \le i<j \le n-1 ; \, n-m+1 \le k \le
 n-1) \quad. \label{geng2c}
\end{align}
\end{subequations}
The sets (\ref{geng2b}) and (\ref{geng2c}) span Euclidean algebras
$\mathsf{eu}(n-m-1)$ and $\mathsf{eu}(m-1)$, respectively;
furthermore, both algebras are ideals in the Lie algebra $\hat{g}_2$,
as follows from three facts: 1.) the adjoint action of $L_{01}$ maps
all $U_{\mu}$ into $-U_{\mu}$, $\mu= 2, \ldots n-1$, see
(\ref{cond3a}); 2.) the adjoint action of $L_{01}$ annihilates all
$L_{\mu\nu}$, $2 \le \mu < \nu \le n-1$, see (\ref{cond3b}); and 3.)
both sets (\ref{geng2b}) and (\ref{geng2c}) commute, see
(\ref{cond3c}, \ref{cond3d}, \ref{cond3e}). $\hat{g}_2$ therefore
contains a direct sum of two commuting Euclidean algebras
$\mathsf{eu}(n-m-1)$ and $\mathsf{eu}(m-1)$ such that this direct sum
is an ideal in the full algebra, in which the ''conformal'' dilation
generator $L_{01}$ acts nontrivially only on the Euclidean
''translation'' generators $U_{\mu}$. The group $G_2$ is then obtained
upon exponentiation of this Lie algebra as
\begin{equation}
\label{pp3fo40}
 G_2 \simeq \left[\EE(n-m-1)\otimes \EE(m-1) \right] \odot
 \RR^{\times}_+\quad ,
\end{equation}
where the group $\RR^{\times}_+$ is defined in formula
(\ref{confgen}). The first Euclidean group $\EE(n-m-1)$ acts only on
those dimensions which are linearly independent from the lattice
dimensions; these dimensions are not involved in the compactification
process and give rise, together with the basis vector $\v{e}_-$, to
the $\RR^{n-m}$-factor in the compactified spacetime. On the other
hand, the second Euclidean group $\EE(m-1)$ acts on the remaining
$(m-1)$ spacelike lattice dimensions spanned by ${\bf
e}_{n-m+1},\ldots,{\bf e}_{n-1}$. The notation in eq. (\ref{pp3fo40})
indicates that the direct product of these groups is normal in the
full group $G_2$, hence the semidirect product with $\RR_+^{\times}$.

So far we have identified the group $G_2$ that preserves the
$\RR$-linear envelope of the lattice vectors. We now implement step
two and replace this $\RR$-linear span by the original $\ZZ$-linear
span, which gives the original lattice $\mathsf{lat}$. The subgroup
$G_{\mathsf{lat}}$ of the Lorentz group we seek is therefore a proper
subgroup of $G_2$ which preserves the lattice points in the sense of
$G_{\mathsf{lat}}\, \mathsf{lat} \subset \mathsf{lat}$. It is clear
that in the process of this reduction the second of the Euclidean
groups, $\EE(m-1)$, now must be replaced by a {\it discrete} group $D$
(which is called {\it maximal point group} in crystallography
\cite{Cornwell1,Cornwell2}). It is also clear that the first Euclidean
group $\EE(n-m-1)$ remains unchanged, since it is not affected by the
replacement $\RR \rightarrow \ZZ$. Finally, the group
$\RR_+^{\times}$, which acts on the lightlike lattice vector ${\bf
e}_+$ according to matrices (\ref{confgen}), must be replaced by a
discrete set of transformations which we now examine:

The group elements $B(\ln a)$, $a\in\RR$, act on ${\bf e}_+$ according
to
\begin{equation}
 B(\ln a)\, {\bf e}_+ = a \cdot {\bf e}_+ \quad.
\label{action1}
\end{equation}
The main point comes now: Since the lattice must be preserved, the
values of $a$ must be restricted to integers, $B(\ln k)$, $k\in\ZZ$,
in order to satisfy $B(\ln k)\, {\bf e}_+ = k \cdot {\bf e}_+ \in
\mathsf{lat}$; we see that this transformation winds the lightlike
circle associated with $\v{e}_+$ $k$ times around itself. Although the
original set $B(\ln a)$ was a group, the set $B(\ln k)$ involving {\it
integers} clearly is no longer a group, but only a {\it semi}group,
since the operation $B(\ln k)$, $k \neq 1$, has no inverse in this
set. This semigroup is isomorphic to the set $(\NN_+,\cdot)$ of
positive integers with multiplicative composition $(z,z')\mapsto
z\cdot z'$, and $1$ as unit. Clearly, $B(\ln k)$ is still an
invertible element of the original Lorentz group $SO_0(1,n-1)$;
however, it has no inverse in $G_{\mathsf{lat}}$, since $B(\ln k)^{-1}
= B(\ln \frac{1}{k})$ is not lattice-preserving, as it maps ${\bf
e}_{+}\mapsto \frac{1}{k} \cdot {\bf e}_{+} \not \in \mathsf{lat}$ ,
for $k>1$.

This analysis can be extended to include the full Lorentz group
$O(1,n-1)$; in this case we find, following the line of arguments
after eq. (\ref{confgen}), that the semigroup $(\NN_+, \cdot)$ is
extended to the semigroup $(\ZZ^{\times}, \cdot)$ of nonzero integers
with multiplication as composition.

In contrast to the semigroup structure that arises in the transition
$\RR^{\times} \rightarrow \ZZ^{\times}$, the discrete subgroup $D$
which acts on the spacelike lattice vectors only is indeed a group, as
follows from theorem [\ref{Bedingung}], since the sub-lattice spanned
by the spacelike lattice vectors satisfies the conditions of this
theorem.

The [semi-]group [$eG_{\mathsf{lat}}$] $G_{\mathsf{lat}}$ then has the
structure
\begin{subequations}
\label{eglat}
\begin{align}
 G_{\mathsf{lat}} & \simeq [\EE(n-m-1) \otimes D ] \quad ,
 \label{eglata} \\
 eG_{\mathsf{lat}} & \simeq [\EE(n-m-1) \otimes D ] \odot
 (\ZZ^{\times},\cdot) \quad. \label{eglatb}
\end{align}
\end{subequations}
Comparison with (\ref{Norm1}, \ref{Norm1}') shows that second line
(\ref{eglatb}) contains the extension of the "ordinary" normalizer
$N(\Gamma)$ by a semigroup $(\ZZ^{\times},\cdot)$, which is what we
wanted to show.

We finish our arguments by giving the expressions for the isometry
group $I(M/\Gamma)$ and its semigroup extension $eI(M/\Gamma)$; they
can be obtained from formulas (\ref{pp3fo3}, \ref{Norm1},
\ref{Norm1}', \ref{eglat}):
\begin{equation}
\label{Exten1}
\begin{aligned}
 I(M/\Gamma) & = \bigg[\, \RR^n/T_{\mathsf{lat}} \,\bigg] \odot
 G_{\mathsf{lat}} \quad , \\
 eI(M/\Gamma) & = \bigg[\, \RR^n/T_{\mathsf{lat}} \, \bigg] \odot
\bigg[ \, G_{\mathsf{lat}} \odot  (\ZZ^{\times},\cdot) \, \bigg]
\quad.
\end{aligned}
\end{equation}

\section{Summary}

We have shown that the isometry group of identification spaces which
arise as a result of compactifying a certain number of dimensions in a
flat pseudo-Euclidean spacetime $\RR^n_t$ can admit a semigroup
extension, related to the concept of the extended normalizer of the
identification group $\Gamma$ in the group of isometries $I(M)$ of the
covering space $\RR^n_t$. The possibility of such an extension hinges
upon the fact whether the restriction of the metric on the covering
space to the $\RR$-linear envelope of the lattice vectors is Euclidean
or not. In the first case, the compactness of the restricted isometry
group on the envelope obstructs such an extension. On the other hand,
if this restricted metric is pseudo-Euclidean, then nontrivial
extensions of the isometry group of the identification space
$M/\Gamma$ exist. We have explicitly provided an example of such an
extension: In the case of a Lorentzian spacetime compactified over a
lightlike lattice, the extension of the isometry group of the
compactified space $\RR^{n-m}\times T^m$ is provided by a semigroup
which is isomorphic to $(\ZZ^{\times},\cdot)$, i.e., the set of all
nonzero integers with multiplication as composition, and $1$ as unit
element.

\acknowledgements{Hanno Hammer acknowledges support from EPSRC
grant~GR/86300/01.}


\end{document}